\documentclass[preprint,aps,nofootinbib]{revtex4}
\usepackage{}
\usepackage{graphicx}
\usepackage{amsmath}
\usepackage{amsfonts}
\usepackage{amssymb}
\usepackage{color}%
\usepackage{dcolumn}
\usepackage{indentfirst}
\providecommand{\U}[1]{\protect\rule{.1in}{.1in}}

\definecolor{lightgray}{rgb}{.7,.7,.7}

\definecolor{red}{rgb}{1,0,0}

\definecolor{blue}{rgb}{0,0,1}

\definecolor{purple}{rgb}{0.6,0.1,0.7}

\newcommand{\f}{\begin{equation}}
\newcommand{\ff}{\end{equation}}
\newcommand{\fa}{\begin{eqnarray}}
\newcommand{\ffa}{\end{eqnarray}}

\begin{document}
\title{Holographic complexity and action growth in massive gravities}
\author{Wen-Jian Pan $^{1}$}
\email{wjpan_zhgkxy@163.com}
\author{Yong-Chang Huang $^{1}$}
\email{ychuang@bjut.edu.cn}
\affiliation{$^1$ Institute of Theoretical Physics,Beijing University of Technology,Beijing,100124,China}

\begin{abstract}
 In this paper, we investigate the growth rates of action for the anti-de Sitter black holes
 in massive-Einstein gravity models and obtain the universal behaviors of the growth rates of action
 (the rates of holographic complexity) within the ``Wheeler-DeWitt''
 (WDW) patch at the late limit. Furthermore, we find that,
 for the static neutral cases, when the same mass of black holes is given,
 the computational speed of the neutral massive black hole is the same as
 its Einstein gravity counterpart, which is independent with the effect of the graviton mass terms;
 nevertheless, for the static charged cases, when the same mass and
 charge parameters of black holes are given, the growth rates of action for the massive charged
 black holes are always superior to the growth rates of action without graviton mass terms,
which directly shows that the massive charged black holes
as computers on the computational speeds are faster than their Einstein gravity counterparts.

\end{abstract}
\maketitle

\section{Introduction}
In the 1970s, Bekenstein \cite{Bekenstein:1973ur} studied black hole physics
and found that the entropy of a black hole satisfies an area law,
which opens a window to study the holography of gravity. Later,
the holographic principle of gravity was proposed explicitly
in \cite{'tHooft:1993gx,Susskind:1994vu}, which is thought of as an important property of
quantum gravity. The holographic principle asserts
that the dynamical behaviors in quantum field theory
at the boundary can be encoded fully by the degrees of freedom
in a gravitational system in the bulk and vice versa. In particular,
the AdS/CFT correspondence greatly strengthens the insightful belief in modern physics.

Recently, there is an intriguing topic on studying the connection
between gravitational theory in bulk and quantum information theory at the boundary \cite{Harlow:2013tf,
Hartman:2013qma,Susskind:2014rva,Susskind:2016tae,Stanford:2014jda,Roberts:2014isa,Susskind:2014jwa,
Alishahiha:2015rta,Barbon:2015ria,Barbon:2015soa,Dvali:2016lnb,
Chemissany:2016qqq,Ben-Ami:2016qex,Momeni:2016ekm,Momeni:2016qfv,Momeni:2016ira}.
In particular, after reconsidering the proposal in \cite{Susskind:2014rva,Susskind:2016tae},
Susskind and his colleagues proposed a new conjecture \cite{Brown:2015bva,Brown:2015lvg}
stating that the quantum computational complexity of a holographic
boundary state is directly associated with the classical gravitational action in a bulk region
in the ``Wheeler-DeWitt'' (WDW) patch, which is called ``complexity-action
duality'' for short. Precisely, the authors argued that
the rate of quantum complexity for the boundary quantum state is
proportional to the growth rate of the gravitational action on shell
in the bulk region in the WDW patch at the late time approximation.
Then the complexity-action duality can be defined by
\begin{align}\label{Com1}
\mathcal{C}=\frac{S}{\pi\hbar},
\end{align}
where $\mathcal{C}$ is the complexity in the quantum information
theory, whose meaning is that the minimum numbers of quantum gates
are required to produce a certain state from a reference state,
and $S$ is the total classical gravitational action
in the bulk region within the WDW patch.
For a system with a given energy, as already shown in literature \cite{Lloyd},
the growth rate of quantum complexity of the boundary state
should have an upper bound,
\begin{align}\label{Com2}
\frac{d\mathcal{C}}{dt}\leq \frac{2M}{\pi\hbar}.
\end{align}
Putting (\ref{Com1}) into (\ref{Com2}), one can obtain
\begin{align}\label{Com3}
\frac{dS}{dt}\leq 2M.
\end{align}
As asserted by Susskind and his colleagues, a neutral static anti-de Sitter (AdS) black hole
satisfies the holographic bound given by Eq.(\ref{Com3}).
For the case of the neutral static AdS-black hole, the relation in (\ref{Com3})
tells us that how fast the information in the quantum computational
theory is stored can be evaluated by the changed rate of the classical
action in the WDW patch in an AdS-black hole. In order to check
the valid of the conjecture, the authors of \cite{Brown:2015lvg}
had tested the holographic bound with the static shells and shock waves, respectively.
If the conjecture is correct, the growth
rate of action implies that the black holes are not only the
densest hard drives in nature, but also the fastest computers in nature.
Since then, many works on holographic information complexity dual to
the classical action for a black hole have been made in
\cite{Cai:2016xho,Brown:2016wib,Lehner:2016vdi,Couch:2016exn,
Yang:2016awy,Chapman:2016hwi,Huang:2016fks,Carmi:2016wjl}.
In particular, the physical meaning of the change rate of action
is considered further in \cite{Cai:2016xho,Huang:2016fks},
and the universal holographic upper bound argued should be expressed
in terms of taking the values of thermodynamics potentials
at the outer horizon and at the inner horizon, respectively.
And then the viewpoint for the change rates of action
was verified in the various stationary AdS-black hole models \cite{Cai:2016xho,Huang:2016fks}.

 In the previous works, ones have obtained a universal
conclusion that, for a given model with given parameters, a black hole is
the fastest computer in nature. In this paper, we wonder
whether this argument under the same conditions is
also valid for the case of massive gravity or not. As a result,
one of our motivations is to study the universal holographic
behaviors of growth rates of action for massive gravity models
in the WDW patch at the late time limit. And then the second question emerging naturally
is what role they can play exactly for the graviton mass terms
as a novel ingredient introduced in the growth rates of action. Thus, we need to
determine the function of the mass terms in this paper.
Then, according to the prescription \cite{Brown:2015bva,Brown:2015lvg,Cai:2016xho,Huang:2016fks},
we find that, for the static massive AdS-black holes,
the only bulk region between the inner horizon (or the singularity) and
the outer horizon for the charged black hole (or the neutral case) within the WDW patch
contributes to the holographic bound value at the late time limit.
Our growth rates of action for the various massive black holes
share the same forms with their counterparts without the mass terms \cite{Cai:2016xho}.
However, comparing the massive and massless cases,
we find that, when the same mass and charge parameters of
black holes are given, the growth rates of action for the charged
massive black holes always are greater than their counterparts for massless cases, which implies that
the charged massive black holes referred to as computers in nature
are always faster than the corresponding ones appearing in \cite{Cai:2016xho}.

Our paper is organized as follows: In Sec.II, we briefly discuss
and present its charged solution for massive gravity model in a four-dimensional spacetime. In Secs.III and IV,
we derive in detail the universal holographic behaviors of the growth rates of action for
massive gravity models within the WDW patch at the late approximation
and demonstrate what role they can play for the graviton
mass terms in the growth rate of complexity if complexity-action duality is correct.
In Sec.V, under the same conditions, we calculate the relevant behaviors
for the case of the charged Banados-Teitelboim-Zanelli
(BTZ) in 2+1-dimensional spacetime.
In the last section, we will give out conclusions and discussions.
\section{The solution of massive charged black hole}
 Before considering the behaviors of the growth rate of action for a
 four-dimensional Einstein-Maxwell theory in the framework of massive gravity \cite{deRham:2010kj,Vegh:2013sk,Cao:2015cti},
 we first need to present its corresponding black hole solution. Now using
 the conventions of \cite{Brown:2015bva,Brown:2015lvg,Cai:2016xho,Huang:2016fks},
 let us write down the massive Einstein-Maxwell action \footnote{Here,
 the coefficients of both the mass terms and electromagnetic field in our action are different from ones in literature \cite{Vegh:2013sk,Cai:2014znn,Cao:2015cza}.}
 consisting of the Ricci scalar, electromagnetic field, cosmological constant
 term, graviton mass terms, and a surface term \cite{Vegh:2013sk,Cai:2014znn,Cao:2015cza,Pan:2016ztm},
 which can be expressed as
\begin{align}\label{S}
S=\int d^4x\sqrt{-g}[\frac{1}{2\kappa^2}(R-2\Lambda)+\frac{m^2}{\kappa^2}(\alpha_1u_1+\alpha_2u_2)-\frac{1}{16\pi}F^2]+\frac{1}{\kappa^2}\int d^3x\sqrt{-\gamma}K,
\end{align}
where
\begin{eqnarray}
u_1&=&tr\mathcal{K},\\
u_2&=&(tr\mathcal{K})^2-tr(\mathcal{K}^2),
\end{eqnarray}
$\alpha_1$, $\alpha_2$ are negative constants, $\kappa^2=8\pi G$,
$K$ is the trace of the extrinsic curvature, and
the matric $\mathcal{{K^{\mu}}_{\nu}}$ is defined by
$\mathcal{{K^{\mu}}_{\nu}}=\sqrt{g^{\mu\alpha}f_{\alpha\nu}}$.
It tells us that the graviton is allowed to obtain its mass $m$ by the reference
metric coupling the bulk metric to break differemophism symmetry.
Varying the action with respect to the metric field and gauge field,
respectively, one can obtain the corresponding equations
of motion
\begin{align}
R_{\mu\nu}-{1\over2}Rg_{\mu\nu}-\frac{3}{L^2}g_{\mu\nu}+m^2\alpha_1(\mathcal{K_{\mu\nu}}-tr\mathcal{K}g_{\mu\nu})
+m^2\alpha_2[2(tr\mathcal{K})\mathcal{K}_{\mu\nu}-2{\mathcal{K}_{\mu}}^{\alpha}\mathcal{K}_{\alpha\nu}]\nonumber\\
-m^2\alpha_2 g_{\mu\nu}[(tr\mathcal{K})^2-tr(\mathcal{K}^2)]=2G(F_{\mu\alpha}{F_{\nu}}^{\alpha}-\frac{1}{4}g_{\mu\nu}F^2),\label{EOMEMG}\\
\nabla_{\mu}F^{\mu\nu}=0\label{EOMEE}.
\end{align}
In order to solve these equations, one can assume that the background spacetime metric field,
and gauge field $A_{\mu}$ have correspondingly the following forms,
respectively:
\begin{align}
ds^2&=-f(r)dt^2+\frac{dr^2}{f(r)}+r^2d\theta^2+r^2\sin^2\theta d\varphi^2,\\
A_{\mu}&=(A_t,0,0,0).
\end{align}
Here following the ansatz in \cite{Cai:2014znn}, the reference metric without dynamical behavior
is chosen as $f_{\mu\nu}=diag(0,0,1, \sin^2\theta)$,\footnote{In the coordinates $(t,r,\theta,\varphi)$,
one can derive the reference metric from $f_{\mu\nu}=\eta_{ab}\partial_{\mu}\phi^a\partial_{\nu}\phi^b$ via taking the $\phi^a$ Stuckelberg fields as
$\phi^0=C,\ \ \phi^1=\sin\theta\cos\varphi, \ \ \phi^2=\sin\theta\sin\varphi$, and $\phi^3=\cos\theta$.} which typically keeps
the differemophism symmetry for coordinates $(t,r)$ intact,
but breaks it in angular directions.
And then one can straightforwardly solve Eqs.(\ref{EOMEMG}) and (\ref{EOMEE}),
and easily find that the corresponding solutions can be presented in the following forms:
\begin{align}
f(r)&=1-\frac{2GM}{r}+\frac{GQ^2}{r^2}+\frac{r^2}{L^2}+m^2\alpha_1r+2m^2\alpha_2\label{SoluMEM}\\
A_t&=-\frac{Q}{r},\label{SoluEM}
\end{align}
where $M$ and $Q$ are the mass and total charge parameters of black hole, respectively.
In the next two sections, for the static black holes including
the neutral and charged two cases in massive gravity theories,
we focus on deriving in detail the behaviors of the growth rates of action within the WDW patch
at the late time limit and understanding the roles of the mass terms in the growth rates of classical action.
\section{The action growth for the neutral black hole}
Now, following the standard procedures in \cite{Brown:2015bva,Brown:2015lvg,Cai:2016xho},
we are interested in the change rate of action in the neutral massive
Schwarzschild-AdS black hole in order to measure how fast the quantum
information can be stored by holographic dual. For this case, the above spacetime
background solution in (\ref{SoluMEM}) naturally becomes the neutral one
with the charge $Q=0$, which turns out to be
\begin{align}
f(r)&=1-\frac{2GM}{r}+\frac{r^2}{L^2}+m^2\alpha_1r+2m^2\alpha_2.\label{SoluMEM2}
\end{align}
In order to calculate the growth rate of total action, by virtue of the action in (\ref{S}),
we can derive such rate through combining the growth rate of the bulk action with
one of the York-Gibbons-Hawking (YGH) surface term. Along with the method
\cite{Brown:2015bva,Brown:2015lvg,Cai:2016xho},
we first determine out
the growth rate of bulk action by integrating over the region within
the WDW patch in the late time approximation.
From the equation of motion (\ref{EOMEMG}), we have
\begin{align}
R=-\frac{12}{L^2}-\frac{6m^2\alpha_1}{r}-\frac{4m^2\alpha_2}{r^2}.\label{RS}
\end{align}
With the use of Eq.(\ref{RS}) we can obtain the growth rate of bulk action,
\begin{align}
\frac{dS_{MEG}}{dt}&=4\pi\int_0^{r_+}drr^2[\frac{1}{2\kappa^2}(R+\frac{6}{L^2})+\frac{m^2}{\kappa^2}(\alpha_1u_1+\alpha_2u_2)]\nonumber\\
&=-\frac{4\pi}{\kappa^2}\int_0^{r_+}dr[\frac{3r^2}{L^2}+m^2\alpha_1r]\nonumber\\
&=-\frac{4\pi}{\kappa^2}[\frac{r_+^3}{L^2}+\frac{m^2\alpha_1}{2}r^2_+].
\end{align}
Now we turn to considering the contribution from the York-Gibbons-Hawking (YGH) surface term in the same limit.
From the definition of the extrinsic curvature, we have
\begin{align}
K=\frac{f_c^{\prime}}{2\sqrt{f_c}}+\frac{2\sqrt{f_c}}{r_c}.
\end{align}
Thus, the contribution from the surface term can be expressed as
\begin{align}
\frac{dS_{YGH}}{dt} &=\frac{4\pi}{\kappa^2}[\frac{r_c^2f_c^{\prime}}{2}+2r_cf_c]_0^{r_+}\nonumber\\
&=\frac{4\pi}{\kappa^2}[\frac{r_+^3}{L^2}+\frac{m^2\alpha_1}{2}r^2_++4GM].
\end{align}
Finally, the growth rate of total action is given by
\begin{align}\label{NSR}
\frac{dS}{dt}&=\frac{dS_{MEG}}{dt}+\frac{dS_{YGH}}{dt}\nonumber\\
&=2M.
\end{align}
The calculating outcome is identical with one of the neutral AdS-black hole \cite{Brown:2015lvg,Cai:2016xho}.
Under the late time approximation, due to the region outside the outer horizon
stationary, it does not contribute to the growth rate of total action. In other words,
the rate relies only on the bulk region between the singularity and the outer horizon.
For the neutral black hole in the massive gravity theory, in light of the complexity-action
duality stated above in the Introduction, its growth rate of holographic complexity can still take the form,
\begin{align}
\frac{d\mathcal{C}}{dt}=\frac{2M}{\pi\hbar}.
\end{align}
This expression indicates that the neutral massive black hole can be still viewed as
the fastest computer in nature.
\section{the action growth for the charged black hole}
In this section, we continuously discuss the growth behavior of the relevant action
for the massive Einstein-Maxwell gravity in a parallel way. Since the massive charged
black hole shares the similar Penrose diagram with the charged black hole (or RN black hole),
in light of the standard procedure \cite{Cai:2016xho}, we can treat it like RN black hole done in literature.
To do this, first of all, the inner horizon $r_-$ and the outer horizon $r_+$ are determined
by solving the solution $f(r_\pm)=0$, which give rise to the following relations:
\begin{align}
M&=\frac{1}{2G}[(r_++r_-)(1+2m^2\alpha_2)+\frac{1}{L^2}(r_++r_-)(r^2_++r^2_-)+m^2\alpha_1(r^2_++r_+r_-+r^2_-)],\label{mass}\\
Q^2&=\frac{r_+r_-}{G}[1+2m^2\alpha_2+\frac{1}{L^2}(r^2_++r_+r_-+r^2_-)+m^2\alpha_1(r_++r_-)]\label{charge}.
\end{align}
With the use of Eq.(\ref{RS}), one then straightforwardly calculates the growth rate of the bulk action, which turns out to be
\begin{align}\label{MEMGA}
\frac{dS_{MEMG}}{dt}=-\frac{1}{2G}[\frac{1}{L^2}(r_+^3-r_-^3)+\frac{m^2\alpha_1}{2}(r_+^2-r_-^2)]-\frac{Q^2}{2}(\frac{1}{r_+}-\frac{1}{r_-}),
\end{align}
whilst, in the same conditions, the contribution from the surface term
for this case gives the following relation:
\begin{align}
\frac{dS_{YGH}}{dt}=-\frac{Q^2}{2}(\frac{1}{r_+}-\frac{1}{r_-})+\frac{1}{2G}[\frac{1}{L^2}(r_+^3-r_-^3)
+\frac{m^2\alpha_1}{2}(r_+^2-r_-^2)].\label{SurT2}
\end{align}
Together the above relation (\ref{SurT2}) with (\ref{MEMGA}), they directly produce
\begin{align}\label{SGR}
\frac{dS}{dt}&=\frac{dS_{MEMG}}{dt}+\frac{dS_{YGH}}{dt}\nonumber\\
&=-Q^2(\frac{1}{r_+}-\frac{1}{r_-})\nonumber\\
&=Q^2(\frac{1}{r_-}-\frac{1}{r_+}).
\end{align}
If we define the chemical potential $\mu_-=\frac{Q}{r_-}$
at the inner horizon and $\mu_+=\frac{Q}{r_+}$ at the outer
horizon, then one can rewrite the relation (\ref{SGR}) as
\begin{align}\label{SRN}
\frac{dS}{dt}=(M-\mu_+Q)-(M-\mu_-Q).
\end{align}
This result on the form is the same as that for the charged black hole
in literature \cite{Cai:2016xho}. If we take the neutral limit, namely,
the charge $Q \to 0$, then, with the use of Eqs.(\ref{mass}) and (\ref{charge})
as well as the definition of the chemical potential, it is easy to find that $\mu_+Q\to 0$ and $\mu_-Q\to 2M$,
so that the rate of change of total action in Eq.(\ref{SRN}) naturally reduces to Eq.(\ref{NSR}).
For the charged case, the holographic complexity can be written as
\begin{align}\label{SCMRN}
\frac{d\mathcal{C}}{dt}=\frac{1}{\pi\hbar}[(M-\mu_+Q)-(M-\mu_-Q)].
\end{align}
Next, we would like to determine the role of
the graviton mass terms. To do this, we attempts to
compare the known result from the charged black hole for the massless case \cite{Cai:2016xho} and
our result with the massive terms. If $\alpha_1$ and $\alpha_2$ are taken as zeros, then the massive charged
black hole solution in (\ref{SoluMEM}) reduces naturally to
a four-dimensional charged black hole geometric solution as shown in \cite{Cai:2016xho}
\begin{align}
\hat{f}(r)=1-\frac{2GM}{r}+\frac{GQ^2}{r^2}+\frac{r^2}{L^2}.\label{RNSo}
\end{align}
Hence, from Eqs. (\ref{SoluMEM}) and (\ref{RNSo}), one easily obtains
\begin{align}\label{dff}
f(r)-\hat{f}(r)=m^2\alpha_1r+2m^2\alpha_2<0,
\end{align}
here we have used the fact that $\alpha_1$, $\alpha_2$ are negative constants.
Correspondingly, the inner horizon $\hat{r}_-$ and the outer horizon $\hat{r}_+$ are determined
by $\hat{f}(\hat{r}_\pm)=0$. As a result,
the relation in (\ref{dff}) implies
\begin{align}
r_-<\hat{r}_-,\ \ \ r_+>\hat{r}_+.
\end{align}
This means that when the mass $M$ and charge $Q$ parameters of black holes are given, Eq.(\ref{SGR}) directly indicates
\begin{align}\label{css}
\frac{dS}{dt}>\frac{d\hat{S}}{dt}.
\end{align}
Note that we have denoted those physical quantities from \cite{Cai:2016xho} with the symbol ``hat''.
Eq.(\ref{css}) directly tells us an interesting conclusion that the massive charged black hole
viewed as a computer in our setup is faster than the one in \cite{Cai:2016xho},
if the complexity-action duality is valid.
\section{Charged BTZ black hole in massive gravity}
Now we turn to considering the case of a charged Banados-Teitelboim-Zanelli
(BTZ) in 2+1-dimensional spacetime \cite{Cai:2016xho,Martinez:1999qi,Clement:1995zt,Cadoni:2007ck}.
The action for a charged-BTZ-AdS black hole can be generalized to involve
the graviton mass terms in the context of massive gravity \cite{Prasia:2016esx,Hendi:2016pvx}, which is presented as
\begin{align}\label{S2}
S=\int d^3x\sqrt{-g}[\frac{1}{2\kappa^2}(R-2\Lambda)+\frac{m^2}{\kappa^2}\alpha_1u_1-\frac{1}{4}F^2]+\frac{1}{\kappa^2}\int d^2x\sqrt{-\gamma}K,
\end{align}
where the reference metric is $f_{\mu\nu}=diag(0,0,1)$.
Varying the action with respect to metric field and gauge field, respectively,
one can easily obtain the following equations of motion:
\begin{align}
R_{\mu\nu}-{1\over2}Rg_{\mu\nu}-\frac{1}{L^2}g_{\mu\nu}+m^2\alpha_1(\mathcal{K_{\mu\nu}}-tr\mathcal{K}g_{\mu\nu})
&=8\pi G(F_{\mu\alpha}{F_{\nu}}^{\alpha}-\frac{1}{4}g_{\mu\nu}F^2),\label{EOMEMG2}\\
\nabla_{\mu}F^{\mu\nu}&=0\label{EOMEE2}.
\end{align}
Suppose that the geometric manifold and the gauge field can take the following forms:
\begin{align}
ds^2&=-f(r)dt^2+\frac{dr^2}{f(r)}+r^2d\theta^2,\\
A_{\mu}&=(A_t,0,0).
\end{align}
Then solving Eqs.(\ref{EOMEMG2}) and (\ref{EOMEE2}), we follow
the usual convention $8G=1$ in \cite{Cai:2016xho} and find that
the solutions can be expressed as
\begin{align}
f(r)&=\frac{r^2}{L^2}+2m^2\alpha_1r-\pi Q^2\ln\frac{r}{L}-M,\\
A_t&=-Q \ln\frac{r}{L}.
\end{align}
The inner horizon $r_-$ and outer horizon $r_+$ are determined by the relation $f(r_\pm)=0$,
such that we can express the mass and charge relations in terms of $r_\pm$, which turn out to be
\begin{align}
Q^2&=\frac{r^2_+-r^2_-}{\pi L^2\ln\frac{r_+}{r_-}}+\frac{2m^2\alpha_1(r_+-r_-)}{\pi\ln\frac{r_+}{r_-}},\\
M&=\frac{1}{L^2\ln\frac{r_+}{r_-}}[r^2_-\ln\frac{r_+}{L}-r^2_+\ln\frac{r_-}{L}]+\frac{2m^2\alpha_1}{\ln\frac{r_+}{r_-}}(r_-\ln\frac{r_+}{L}-r_+\ln\frac{r_-}{L}).
\end{align}
Analogy with the charged massive-AdS black hole, then one can simply compute
the rate of total action within the WDW patch at the late limit,
\begin{align}\label{SRNBTZ}
\frac{dS}{dt}&=2\pi Q^2[\ln\frac{r_+}{L}-\ln\frac{r_-}{L}]\nonumber\\
&=\mu_-Q-\mu_+Q\nonumber\\
&=(M-\mu_+Q)-(M-\mu_-Q),
\end{align}
where the chemical potential at each horizon is defined by
\begin{align}
\mu_+&=-2\pi Q\ln\frac{r_+}{L},\\
\mu_-&=-2\pi Q\ln\frac{r_-}{L}.
\end{align}
The holographic complexity thus can still be presented as
\begin{align}
\frac{d\mathcal{C}}{dt}=\frac{1}{\pi\hbar}[(M-\mu_+Q)-(M-\mu_-Q)].
\end{align}
One can easily check that when $\alpha_1=0$, all the relations above
naturally reduce to those of a charged-BTZ-AdS black hole as stated by \cite{Cai:2016xho}.
Next, if taking the neutral charge limit $Q\to0$, then $\mu_+Q\to 0$,
while $\mu_-Q\to 2M$, this means that
\begin{align}\label{SRNBTZ2}
\frac{dS}{dt}=2M.
\end{align}
Similarly, taking the graviton mass terms into account, we can show that
the growth rate of action for massive charged BTZ black hole is superior to that
for the corresponding BTZ black hole. To be specific, we explicitly present
the results for charged BTZ black hole in \cite{Cai:2016xho}, in which
the geometric solution is written as
\begin{align}
\hat{f}(r)=\frac{r^2}{L^2}-\pi Q^2\ln\frac{r}{L}-M,
\end{align}
such that
\begin{align}
f(r)-\hat{f}(r)=2m^2\alpha_1r<0,
\end{align}
this formula implies that their inner horizons and outer horizons should satisfy the following relations, respectively:
\begin{align}
r_-<\hat{r}_-,\ \ \ r_+>\hat{r}_+.
\end{align}
Therefore, from Eq.(\ref{SRNBTZ}), one can get
\begin{align}
\frac{dS}{dt}-\frac{d\hat{S}}{dt}>0.
\end{align}
Here, we still obtain an interesting conclusion that the rate of holographic complexity
in our case exceeds the rate of holographic complexity in literature \cite{Cai:2016xho},
which means that the massive BTZ black hole as a computer is still faster than
that in \cite{Cai:2016xho}.
\section{conclusions and discussions}
In this paper, inspired by the recent works \cite{Brown:2015bva,
Brown:2015lvg,Cai:2016xho}, we have explored the universal behaviors
of the growth rates of action \footnote{We have checked that, under the same conditions, the current
calculating outcomes for the growth rates of action in massive gravity
are consistent with the results by using the procedure in
\cite{Lehner:2016vdi,Chapman:2016hwi}. We thank the anonymous referee for raising this
issue with us.} for the static massive-AdS black holes
within the WDW patch at the late time approximation, which should be
dual to the universal behaviors of the growth rates of
the holographic bound states at the AdS boundary.
Along with the procedure in \cite{Cai:2016xho}, we find that
the change rate of action for the neutral black hole
in the massive-Einstein gravity is just equal to $2M$;
whilst the rates for the charged (RN and BTZ) AdS-black holes
in the massive Einstein gravity still depend only on the difference of
the quantity $M-\mu Q$ taking values at the outer horizon
and inner horizon, respectively, which share the same structure
with their Einstein gravity counterparts \cite{Cai:2016xho}. These results
thus have further strengthened the viewpoints of the universal bounds
for the growth rates of action proposed in \cite{Cai:2016xho}.
In addition, our outcomes have been shown that
the massive black holes are referred to as the fastest computers in nature.
If the complexity-action duality is correct, our results mean
that the holographic complexity grows linearly with time.

From our calculations for the growth rates of action in massive black holes,
although our results share the same expressions with the corresponding
ones known in \cite{Cai:2016xho}, they have been shown that the effects of
the graviton mass terms can stimulate their computational speeds of
the charged black hole computers. In particular, we have found that,
for the static neutral cases, when the same mass is given, the massive
AdS-black hole computer is as fast as the massless AdS-black hole computer,
which is independent with the effects of the graviton mass terms; nevertheless,
for the static charged cases, when the same mass
and charge parameters of black holes are given, the growth rates of action
in the massive gravity are always superior to the
growth rates of action without graviton mass terms, which directly shows
that the massive charged black holes as computers on
the computational speeds are faster than the ones without
taking the mass terms.

 Note that, for a system with given parameters(the integral constants), a universal
conclusion obtained is that black holes in the current models
including massive gravity are the fastest computers in nature,
since their rates of dual complexity saturate the Lloud bound. However,
this universal conclusion can not avoid the differences
in various gravity models with given parameters.
For example, for the neutral static case with a given mass parameter,
due to the Gauss-Bonnet term, the growth rate of action for
Gauss-Bonnet gravity is smaller than its Einstein gravity counterpart \cite{Cai:2016xho}.
Similarly, the striking fact that the computational
speed of the massive charged black holes is faster than the one
in massless cases, originates from the effect of graviton mass terms.
But it is worthy to note that this does not violate the universal conclusion,
since the Lloud bound is still kept intact. Here, comparing the massive cases
and massless cases, we want to distinguish the rates of holographic
complexity for massive black holes from their Einstein gravity counterparts such that
the second motivation can be determined. As a consequence,
the outcomes that the rates of holographic complexity for massive charged black holes
in our paper are larger than the ones for massless cases are interesting conclusions.

\begin{acknowledgments}
 Wen-Jian Pan would like to thank Shao-Jiang Wang, Jian-Pin Wu,
 Yu Tian, Min-Yong Guo, Xiao-Mei Kuang and Zhuo-Yu Xuan
 for useful discussions. This work is supported by the National Natural Science
 Foundation of China (No. 11275017 and No. 11173028).
\end{acknowledgments}

\end{document}